\documentclass[%
reprint,nofootinbib,
superscriptaddress,
 amsmath,amssymb,
 aps,prd,longbibliography, 
]{revtex4-1}

\usepackage[utf8]{inputenc}
\usepackage[T1]{fontenc}
\usepackage{mathptmx}
\usepackage{subfig}
\usepackage{tabularx}
\usepackage[english]{babel}
\usepackage[dvipsnames]{xcolor}
\usepackage[breaklinks,pdftex,pdfpagelabels,bookmarks]{hyperref}
\hypersetup{colorlinks=true,linktocpage=true,pdfstartpage=3,pdfstartview=FitH,breaklinks=true,pdfpagemode=UseNone, pageanchor=true,pdfpagemode=UseOutlines,plainpages=false,bookmarksnumbered,bookmarksopen=true,bookmarksopenlevel=1,hypertexnames=true,pdfhighlight=/0,urlcolor=Brown,linkcolor=MidnightBlue!60!Black,citecolor=PineGreen}
\usepackage[version=4]{mhchem}
\usepackage{siunitx}
\usepackage{comment}
\usepackage{graphicx}% Include figure files
\usepackage{caption}
\usepackage{dcolumn}% Align table columns on decimal point
\usepackage{bm}% bold math
\newcommand{\ped}[1]{_{\text{#1}}}

\begin{document}

\preprint{APS/123-QED}

\title{Electrodynamics of highly spin-polarized tunnel Josephson junctions}% Force line breaks with \\
%\thanks{A footnote to the article title}%

\author{H. G. Ahmad}
\affiliation{Dipartimento di Fisica E. Pancini$,$ Università degli Studi di Napoli Federico II$,$ Monte S. Angelo$,$ via Cinthia$,$ I-80126 Napoli$,$ Italy}
\affiliation{CNR-SPIN$,$ UOS Napoli$,$ Monte S. Angelo$,$ via Cinthia$,$ I-80126 Napoli$,$ Italy}
\email{halimagiovanna.ahmad@unina.it}
\author{R. Caruso}
\affiliation{Dipartimento di Fisica E. Pancini$,$ Università degli Studi di Napoli Federico II$,$ Monte S. Angelo$,$ via Cinthia$,$ I-80126 Napoli$,$ Italy}
\affiliation{CNR-SPIN$,$ UOS Napoli$,$ Monte S. Angelo$,$ via Cinthia$,$ I-80126 Napoli$,$ Italy}
\affiliation{SeeQC-eu$,$ via dei Due Macelli 66$,$ I-00187 Roma$,$ Italy}
\author{A. Pal}
\affiliation{Department of Materials Science and Metallurgy$,$ University of Cambridge$,$ 27 Charles Babbage Road$,$ Cambridge CB3 0FS$,$ United Kingdom}
\affiliation{Department of Metallurgical Engineering and Materials Science$,$ IIT Bombay$,$ Mumbai$,$ 400076$,$ India}
\author{G. Rotoli}
\affiliation{Università della Campania "Luigi Vanvitelli"$,$ Dipartimento di Ingegneria$,$ via Roma 29$,$ 81031 Aversa (CE)$,$ Italy}
\author{G. P. Pepe}
\affiliation{Dipartimento di Fisica E. Pancini$,$ Università degli Studi di Napoli Federico II$,$ Monte S. Angelo$,$ via Cinthia$,$ I-80126 Napoli$,$ Italy}
\affiliation{CNR-SPIN$,$ UOS Napoli$,$ Monte S. Angelo$,$ via Cinthia$,$ I-80126 Napoli$,$ Italy}
\author{M. G. Blamire}
\affiliation{Department of Materials Science and Metallurgy$,$ University of Cambridge$,$ 27 Charles Babbage Road$,$ Cambridge CB3 0FS$,$ United Kingdom}
\author{F. Tafuri}
\affiliation{Dipartimento di Fisica E. Pancini$,$ Università degli Studi di Napoli Federico II$,$ Monte S. Angelo$,$ via Cinthia$,$ I-80126 Napoli$,$ Italy}
\affiliation{CNR-SPIN$,$ UOS Napoli$,$ Monte S. Angelo$,$ via Cinthia$,$ I-80126 Napoli$,$ Italy}
\author{D. Massarotti}
\affiliation{CNR-SPIN$,$ UOS Napoli$,$ Monte S. Angelo$,$ via Cinthia$,$ I-80126 Napoli$,$ Italy}
\affiliation{Dipartimento di Ingegneria Elettrica e delle Tecnologie dell’Informazione$,$ Università degli Studi di Napoli Federico II$,$ via Claudio$,$ I-80125 Napoli$,$ Italy}

\begin{abstract}
The continuous development of superconducting electronics is encouraging several studies on hybrid Josephson junctions (JJs) based on superconductor/ferromagnet/superconductor (SFS) heterostructures, as either spintronic devices or switchable elements in quantum and classical circuits. Recent experimental evidence of macroscopic quantum tunneling and of an incomplete $0$-$\pi$ transition in tunnel-ferromagnetic spin-filter JJs could enhance the capabilities of SFS JJs also as active elements. Here, we provide a self-consistent electrodynamic characterization of \ce{NbN}/\ce{GdN}/\ce{NbN} spin-filter JJs as a function of the barrier thickness, disentangling the high-frequency dissipation effects due to the environment from the intrinsic low-frequency dissipation processes. The fitting of the $I-V$ characteristics at $\SI{4.2}{\K}$ and at $\SI{300}{\milli\K}$ by using the Tunnel Junction Microscopic model allows us to determine the subgap resistance $R\ped{sg}$, the quality factor $Q$ and the junction capacitance $C$. These results provide the scaling behavior of the electrodynamic parameters as a function of the barrier thickness, which represents a fundamental step for the feasibility of tunnel-ferromagnetic JJs as active elements in classical and quantum circuits, and are of general interest for tunnel junctions other than conventional SIS JJs.
%\begin{description}
%\item[Usage]
%Secondary publications and information retrieval purposes.
%\item[PACS numbers]
%May be entered using the \verb+\pacs{#1}+ command.
%\item[Structure]
%You may use the \texttt{description} environment to structure your abstract;
%use the optional argument of the \verb+\item+ command to give the category of each item. 
%\end{description}
\end{abstract}

%\pacs{Valid PACS appear here}% PACS, the Physics and Astronomy
%                             % Classification Scheme.
%%\keywords{Suggested keywords}%Use showkeys class option if keyword
%                              %display desired
\maketitle

%\tableofcontents

\section{\label{sec:level1}Introduction}

Ferromagnetic Josephson junctions (SFS JJs) have attracted considerable attention in the emerging fields of superconducting spintronics~\cite{Bergeret2005,Robinson2010,Khaire2010,Eschrig2011,Linder2015} and as quantum and classical devices, since they have been proposed as energy-efficient memories~\cite{Ryazanov2001,Larkin2012,Goldobin2013,Niedzielski2018} and as passive $\pi$ shifters (phase inverters) in quantum circuits~\cite{Ustinov2003,Buzdin2005,Feofanov2010}. However, in standard metallic SFS JJs, the $I\ped c R\ped N$ product is of the order of a few microvolts or less~\cite{Buzdin2005,Robinson2010,Khaire2010}, $I\ped c$ and $R\ped N$ being,  respectively, the critical current and the normal state resistance. All these JJs are overdamped and thus characterized by high quasiparticle dissipation~\cite{Buzdin2005,Kato2007,Massarotti2018}. This has hampered the use of ferromagnetic JJs as active switching elements in different classical and quantum circuits, since for such applications it is important to have a rather high $I\ped c R\ped N$ product and low damping~\cite{Kato2007}. Low-dissipative ferromagnetic junctions use an additional insulating layer between one of the superconducting electrodes and the ferromagnetic barrier (SIFS JJs)~\cite{Weides2006,Bannykh2009,Wild2010,Larkin2012} or a ferromagnetic insulator barrier ($\text{SI}\ped{f}\text{S}$ JJs)~\cite{Terzioglu1998,Ioffe1999,Kawabata2006,Kawabata2010,Vasenko2011} and may present key advantages for some applications, thus increasing the overall impact of JJs based on ferromagnetic barriers~\cite{Bannykh2009,Wild2010,Larkin2012,20,21}.

Heterostructures incorporating ferromagnetic insulator tunnel barriers have been theoretically proposed as quantum devices such as \emph{quiet} ferromagnetic flux-qubits, based on anomalous $0$-$\pi$ transitions~\cite{Ioffe1999,Kawabata2006,Kawabata2010}, and as classical devices for digital electronics~\cite{Terzioglu1998} and efficient electron refrigeration~\cite{Kawabata2013refr}. Among the ferromagnetic insulators, \ce{GdN} has been used in superconducting spin valves~\cite{Zhu2016}, switchable JJs based on the interfacial exchange field~\cite{Cascales2019},  and in spin-filter \ce{NbN}/\ce{GdN}/\ce{NbN} JJs, which represent the first $\text{SI}\ped{f}\text{S}$ JJs. Some of their properties have been studied in Refs.~\cite{Senapati2011,28,Pal2014,Massarotti2015,Caruso2019}. The first evidence of macroscopic quantum tunneling (MQT) in ferromagnetic JJs is an indication that spin-filter JJs can be used as active quantum devices~\cite{Massarotti2015}. These JJs are characterized by a thickness-dependent spin polarization because of the splitting in the \ce{GdN} insulator band structure induced by its magnetic exchange energy~\cite{Senapati2011}. This property, together with the nontrivial magnetic structure of the barrier, causes an incomplete $0$-$\pi$ transition for the spin-filter efficiency ($P$) above $90\%$. Such an incomplete $0$-$\pi$ transition could be related to the presence of spin-triplet correlations, with implications for the $0$-$\pi$ technology~\cite{Caruso2019}.
\begin{figure}[t]
	\centering      
	\subfloat[][]{\includegraphics[scale=0.35]{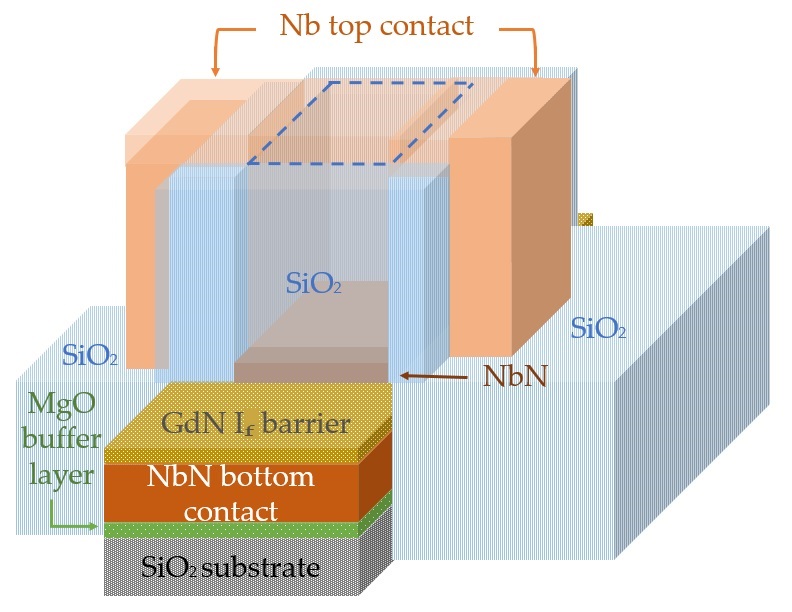}}\qquad
	\subfloat[][]{\includegraphics[width=8.6cm]{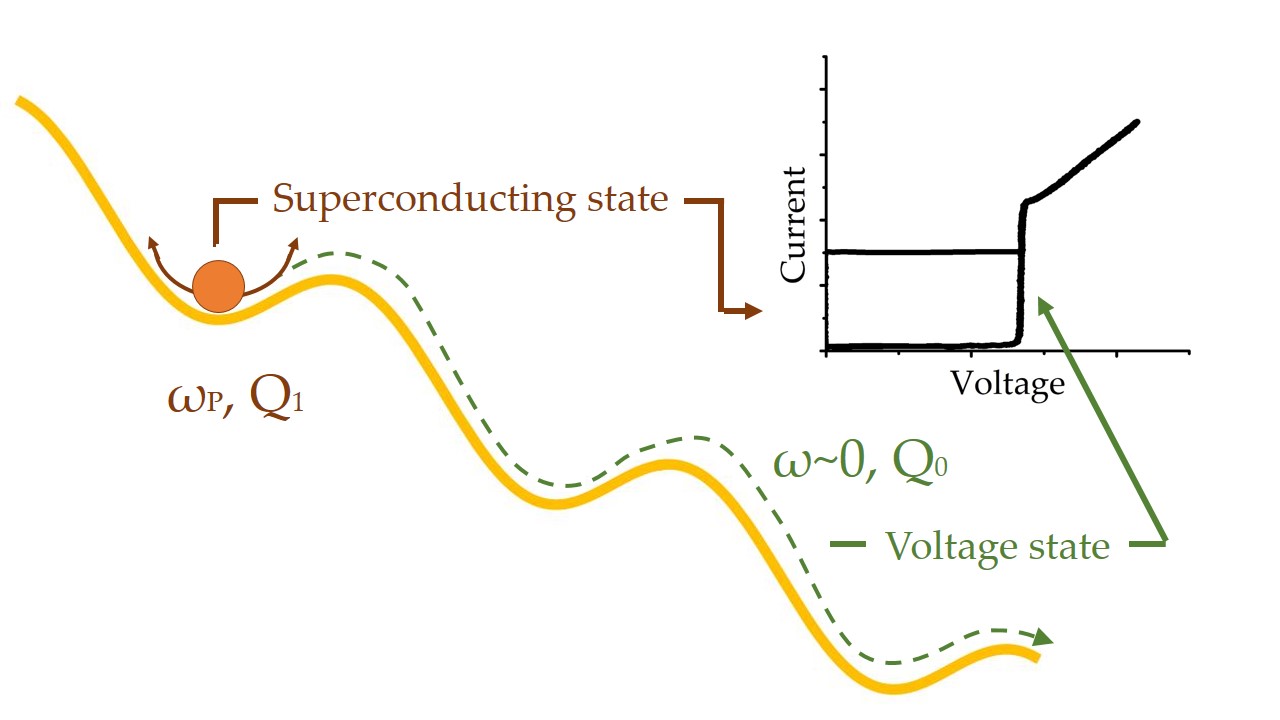}}\\
	\caption[]{In (a): sketch of spin-filter JJs.  The area of the devices is $7\times7\;\si{\micro\m^2}$ (dashed blue window). In (b): washboard potential of a Josephson junction. The brown double arrow represents the oscillating motion at the plasma frequency of the phase particle in the superconducting state. In this regime, the damping is determined by the high-frequency quality factor $Q_1$. The dashed green line represents the steady motion of the particle that rolls down the washboard in the voltage state. In this regime, the damping is determined by the low-frequency quality factor $Q_0$.}
	\label{Fig:1}
\end{figure}

This work aims at providing a self-consistent determination of the electrodynamic parameters in highly spin-polarized \ce{NbN}/\ce{GdN}/\ce{NbN} junctions [Fig.~\ref{Fig:1}~(a)]. For conventional JJs in the underdamped regime and with large $I\ped c$, measurements of Fiske steps have been successfully used to derive the capacitance $C$, while the amplitude of the hysteresis in the $I-V$ curve allows us to estimate the quality factor $Q$ within the resistively and capacitively shunted junction (RCSJ)  model~\cite{Barone1982,Likharev1986}. However, when the junctions fall in the moderately damped regime or are characterized by low values of $I\ped c$ (or critical current density $J\ped c=I\ped c/A$, where $A$ is the cross section), it is more complicated to isolate the effective capacitance and the intrinsic dissipation sources of the junction from contributions due to the environment and the external circuit. Thus, more sophisticated methods are required for the analysis of the dissipation~\cite{Barone1982,Devoret1984,Martinis1987,Martinis1989,Kautz1990}. We use the conventional tunnel junction microscopic (TJM) model to obtain a self-consistent estimation of $C$, $Q$, and the resistance associated with the quasiparticle dynamics $R\ped{sg}$, which are essential to define the electrodynamic properties of devices with $I\ped c$ down to few nanoamperes. The merit of this approach is a comparative analysis of tunnel-ferromagnetic JJs with barrier thickness spanning from $2.5$ to $\SI{4.0}{\nm}$. This allows us to explore substantially quite different transport regimes. If one wants to place the junction in a circuit or to couple it to a cavity~\cite{Devoret2013,Krantz2019}, knowledge of the electrodynamic parameters and how they scale with the barrier thickness is fundamental. Therefore, this study provides a pathway to the engineering of tunnel-ferromagnetic JJs for specific applications.

\section{Methods}
\label{Methods}

Dissipation in a JJ is frequency dependent and the quality factor is given by $Q(\omega)= \omega\ped p R(\omega) C$, where $\omega\ped p=(2eI\ped c/(\hbar C) )^{1/2}$ is the plasma frequency~\cite{Barone1982,Likharev1986}. In terms of the phase dynamics in the tilted washboard potential~\cite{Barone1982}, the phase particle in the supercurrent branch  oscillates in one well of the potential at the plasma frequency $\omega\ped p$, while the voltage state involves steady motion of the phase particle ($\omega\sim 0$)~\cite{Kautz1990}[Fig.~\ref{Fig:1}~(b)]. High-frequency ($\omega\sim\omega\ped P$) dissipation at the switching from the superconducting to the resistive state (see the brown double arrow in Fig.~\ref{Fig:1}~(b)) is determined by the high-frequency damping $Q_1$ and is mainly affected by the environment, i.\,e., the circuit in which the junction is embedded~\cite{Devoret1984,Martinis1987,Martinis1989,Kautz1990}. Low-frequency dissipation in the subgap branch of the $I-V$ curves ($\omega\sim0$) (see the green dashed arrow in Fig.~\ref{Fig:1}~(b)) and the corresponding low-frequency damping $Q_0$ are affected by the intrinsic tunnel resistance, which is set by the subgap resistance $R\ped {sg}$~\cite{53,54,56,Martinis1989,Kautz1990} as
\begin{equation}
	\label{Eq:3}
	Q_0=\omega\ped p C R\ped{sg}.
	\end{equation}
The TJM model provides a complete microscopic description of a JJ, using the tunneling-Hamiltonian formalism~\cite{Barone1982,Likharev1986}, and it is commonly employed for modeling superconducting quantum-interference devices (SQUIDs) and rapid-single-flux-quantum (RSFQ) logic gates and circuits~\cite{43,Odintsov1987,44,45}. This model can describe the subgap branch and the low-frequency electrodynamics of any JJ that shows tunneling conduction, without taking into account the exact expression for the current-phase relation (CPR), which could be nontrivial in the case of unconventional JJs such as the spin-filter JJs analyzed in this work~\cite{Pal2014}. Therefore, it provides a powerful tool to investigate and determine $Q_0$ in junctions far from the underdamped regime and it enables us to isolate the dissipative components coming from the environment. It is particularly relevant since quasiparticle tunneling is a figure of merit in all classical and quantum circuits and has been, in general, a limit for standard SFS JJs.
Measurements down to $\SI{300}{\milli\K}$ of the $I-V$ characteristics are performed by using an evaporation cryostat, while measurements down to $\SI{20}{\milli\K}$ are performed in a wet dilution refrigerator. Customized RC, copper powder filters, and room-temperature electromagnetic interference (EMI) filters guarantee high precision and resolution in the microvolt and nanoampere range. More details on the measurement setup can be found in Refs.~\cite{41,Massarotti2015}, while information regarding the fabrication processes is given in Refs.~\cite{Senapati2011,Blamire2012,Pal2014}. We measure the $I-V$ curves of junctions with different \ce{GdN} thickness $t$ at $\SI{20}{\milli\K}$, $\SI{300}{\milli\K}$, and $\SI{4.2}{\K}$ by current biasing the samples with a triangular waveform at $\SI{11.123}{\hertz}$ and by measuring the voltage across the junction. We extract $I\ped c$ at a voltage value far from the noise detected in the supercurrent branch. The normal resistance $R\ped N$ is calculated with a linear fit above $V\ped g=(\Delta_1+\Delta_2)/e=\SI{3.50}{\milli\V}$, with $\Delta_1$ and $\Delta_2$ being the gap energies of the two superconducting \ce{NbN} electrodes.

\begin{figure}
	\centering
	\includegraphics[width=8.6cm]{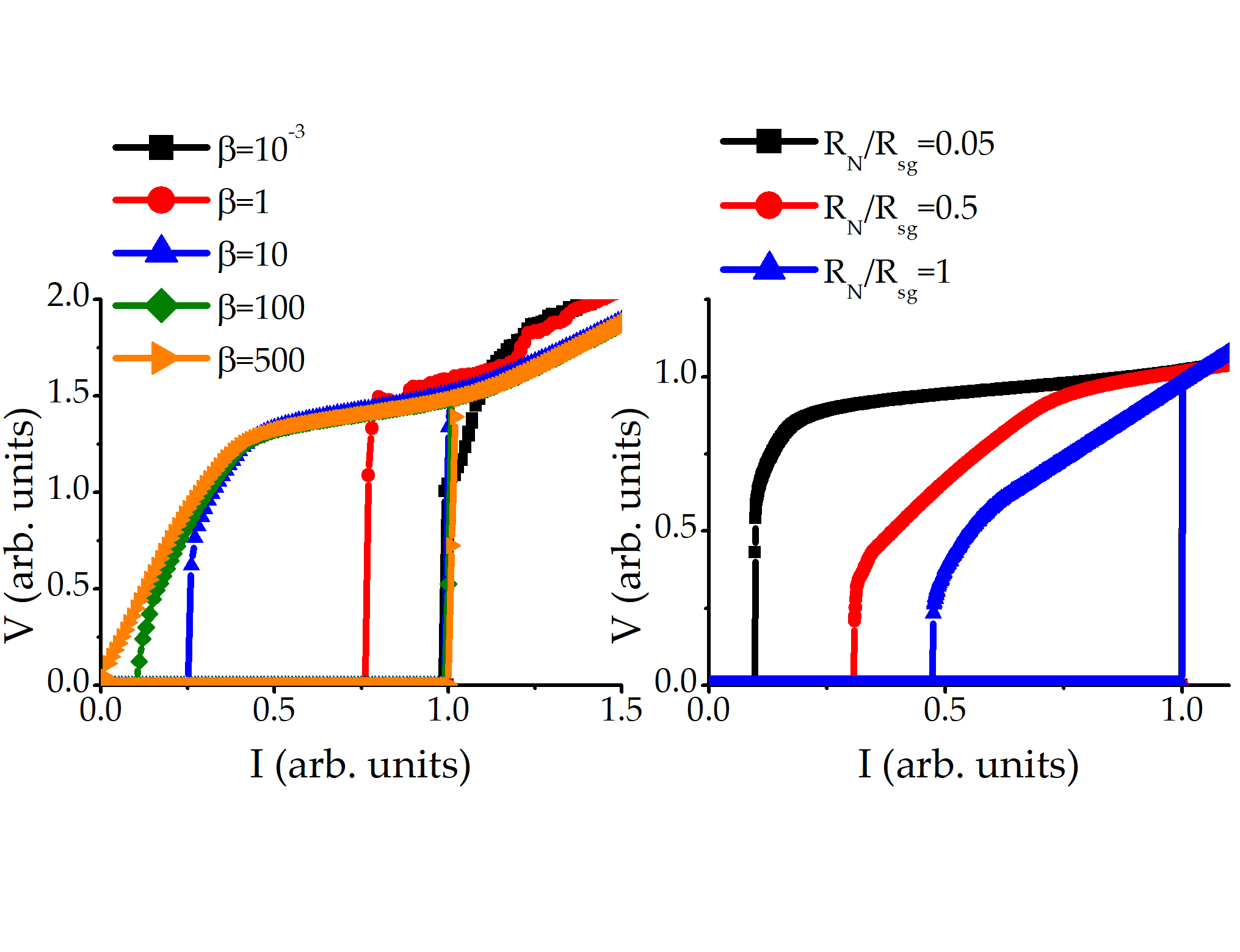}
	\caption[]{$I-V$ curves in normalized units simulated by using \textsc{pscan2}, by fixing $V\ped g=1.4$, $I\ped cR\ped N=1.0$: in a) we fixed $R\ped N/R\ped {sg}=0.1$ and we changed $\beta$; in b) we fixed $\beta=10$ and we changed $R\ped N/R\ped {sg}$.}
	\label{Fig:5}
\end{figure}

TJM simulations are calculated by using \textsc{pscan2}~\cite{url}, a \textsc{python} module optimized to simulate SFQ logic-based superconducting circuits that typically work at $\SI{4.2}{\K}$. One of the subroutines of this software allows to simulate the $I-V$ characteristic of a JJ in electronic circuits with different degrees of complexity ${}^{1}$(The \textsc{pscan2} subroutine calculates time-averaged voltages $V$ across the device as a function of a bias current $I$). $I\ped c$, the Stewart-McCumber parameter $\beta=Q_0^2$, the gap voltage $V\ped g$, the ratio $I\ped cR\ped N/V\ped g$, and the ratio $R\ped N/R\ped{sg}$, $R\ped{sg}$ being the resistance of the subgap branch, are the software parameters that govern the shape of the $I-V$ curves. $I\ped c$ and $V\ped g$ measured directly from the $I-V$ curves in our experimental setup are affected by errors of $1\%$ and $2\%$, respectively, while $R\ped N$ is obtained by fitting the ohmic region of the $I-V$ curves and is affected by an error of $3\%$. Since these values can be obtained with high precision, they can be set as fixed parameters, as well as the ratio $I\ped cR\ped N/V\ped g$. $\beta$ and the ratio $R\ped N/R\ped{sg}$ are the fitting parameters. The Stewart-McCumber parameter modifies the amplitude of the hysteresis in the $I-V$ curve, without affecting the subgap region [Fig.~\ref{Fig:5}~(a)]. The ratio $R\ped N/R\ped{sg}$, instead, modifies both the subgap shape and the hysteresis amplitude [Fig.~\ref{Fig:5}~(b)].

In our simulations, we reproduce the current biasing of a JJ with a current generator in series with the filtered lines of our experimental setup (approximately $\SI{200}{\ohm}$). For each spin-filter JJ with a certain \ce{GdN} thickness $t$, we choose the best-fit parameters $Q_0$ and $R\ped{sg}$ in such a way that the deviations from the experimental curves are minimal. The errors on $Q_0$ and $R\ped{sg}$ represent the range of values that provide a significant overlap between the experimental $I-V$ characteristics and the simulated curves within the TJM model and are of $6\%$ and $10\%$ respectively.

The \ce{GdN} thicknesses in the junctions analyzed in this work range from $2.5$ to $\SI{4.0}{\nm}$, while $P$ ranges from $88\%$ to $98\%$, respectively (Tab.~\ref{Tab:1}), falling in the highly spin-polarized regime. In the special case of spin-polarized systems, $R\ped N$ has to be redefined as the combination of the two resistances associated with the presence of different tunnel conductances for spin-up and spin-down electrons, because of the spin-filtering effect (see the Appendix). The subgap shape in the $I-V$ curves is linked to the quasiparticle dynamics in the junction. The quasiparticle current in a  spin-polarized system has been expressed theoretically and analytically in the case of symmetric spin-filter JJs by taking into account the magnetic nature of the tunnel barrier and the spin-filtering effect~\cite{Bergeret2012}. Simple calculations allow us to verify that the quasiparticle current in these devices has the same expression both in the case of conventional tunnel JJs, i.\,e., for $P=0$ and a magnetic exchange field in the tunnel barrier $h=0$, and in the ideal and extreme situation of perfect spin polarization ($P=100\%$) (see the Appendix). The conventional TJM model does not take into account the magnetic exchange field of the $\text{I}\ped f$ barrier in spin-filter junctions, which can be important in the intermediate regime between these two extreme cases.

The systematic fitting of the $I-V$ curves at $\SI{4.2}{\K}$ as a function of the barrier thickness confirms that the shape of the $I-V$ curves is mostly determined by the standard parameters of the junction ($C$, $R\ped {sg}$, $Q_0$). Further consistency is given by the $I-V$ fitting through the frequency dependent RCSJ model for the junction with the highest $P$, as shown in section~\ref{Results}. Below $\SI{4.2}{\K}$, \ce{NbN}/\ce{GdN}/\ce{NbN} JJs with $P$ up to $98\%$ show an incipient $0$-$\pi$ transition in the $I\ped c(T)$ curves, which can be understood in terms of spin-triplet correlations arising because of the presence of both the spin-filtering effect and a nonuniform magnetic activity in the $\text{I}\ped f$ barrier~\cite{Caruso2019}. Therefore, deviations between the experimental curves and simulations at $\SI{300}{\milli\K}$ can be due to the magnetic nature of the barrier, which the TJM model does not take into account. However, the estimated fitting parameters give an upper bound to $Q_0$ and $R\ped{sg}$, and a term of comparison for possible applications of spin filter JJs at very low temperatures, as discussed in section~\ref{Discussions}.  

\section{Results}
\label{Results}

\begin{figure}[t]
	\centering
	\includegraphics[width=8.6cm]{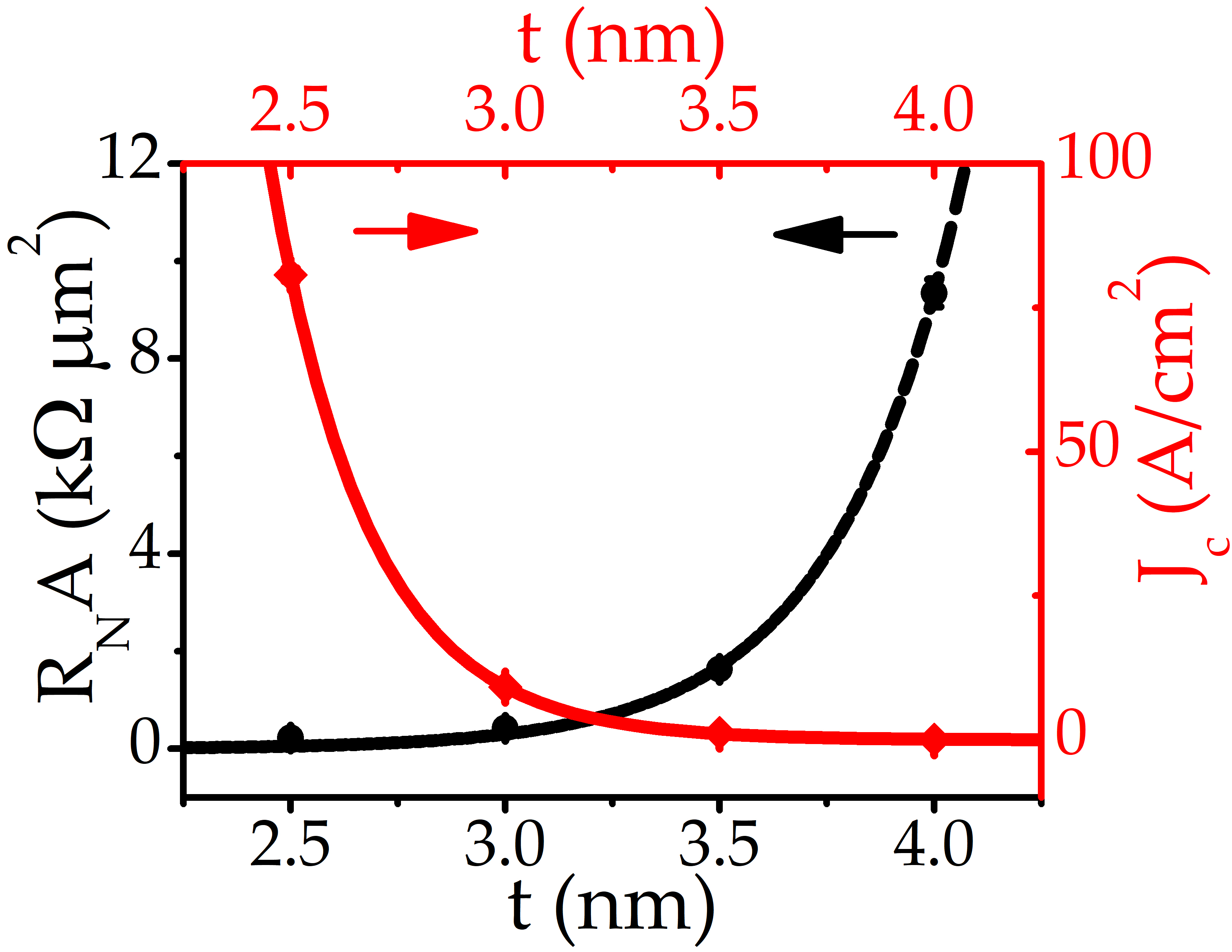}
	\caption[]{In black, $R\ped NA (t)$ product measured at $\SI{300}{\milli\K}$ (black circles) as a function of the barrier thickness along with a fit using Eq.~\ref{Eq:1} (dashed curve). In red, critical current density $J\ped c(t)$ measured at $\SI{300}{\milli\K}$ as a function of the \ce{GdN} thickness $t$ (diamonds) along with an exponential fit (full line). The error bars are of the order of $1\%$ on measured values for $I\ped c$, and of $3\%$ for $R\ped N$.}
	\label{Fig:4}
\end{figure}

As one can observe in Fig.~\ref{Fig:4}, the critical current density $J\ped c(t)$, with cross section $A=\SI{49}{\micro\m^2}$, and the $R\ped NA(t)$ curves at $\SI{300}{\milli\K}$ obey to a typical tunnel behavior, thus confirming the insulating nature of the ballistic \ce{GdN} barrier~\cite{Caruso2019}. $R\ped NA(t)$ exhibits the characteristic exponential thickness dependence:
\begin{equation}
R\ped NA(t)=\frac{2tA}{3\sqrt{4m\ped e\bar{E}}}\left(h/e\right)^2e^{\frac{2t}{\hbar}\sqrt{4m\ped e \bar{E}}},
\label{Eq:1}
\end{equation} 
where $m\ped e$ is the free electron mass, $e$ is the electron charge and $\bar{E}$ is the mean energy-barrier height seen by the charge carriers~\cite{51}.
\begin{table*}
	\centering
	\caption[]{Parameters of the measured spin-filter junctions: thickness $t$, spin-filtering efficiency $P$, characteristic voltage $I\ped c R\ped N$ at $\SI{4.2}{\K}$, $\SI{300}{\milli\K}$ and $\SI{20}{\milli\K}$. The errors on the characteristic voltage are given by a propagation of maximum errors on $I\ped c$ ($1\%$) and $R\ped N$ ($3\%$).}
	\begin{ruledtabular}
		\begin{tabular}{ccccc}
			$t\;(\si{\nm})$ & $P\;(\%)$ & $I\ped cR\ped N@\SI{4.2}{\K}\;(\si{\micro\volt})$ & $I\ped cR\ped N@\SI{300}{\milli\K}\;(\si{\micro\volt})$ & $I\ped cR\ped N@\SI{20}{\milli\K}\;(\si{\micro\volt})$ \\
			\hline
			$2.5$ & $88$ & $156\pm3$ & $179\pm4$ & - \\
			$3.0$ & $93$ & $24.2\pm0.5$ & $38.3\pm0.8$ & $44.0\pm0.9$\\
			$3.5$ & $96$ & $9.9\pm0.2$ &$19.0\pm0.4$ & - \\
			$4.0$ & $98$ & $2.8\pm0.1$ &$5.1\pm0.2$ & $6.1\pm0.2$\\
		\end{tabular}
	\end{ruledtabular}
	\label{Tab:3}
\end{table*}
In Tab.~\ref{Tab:3}, we report $I\ped cR\ped N$ at $\SI{4.2}{\K}$, at $\SI{300}{\milli\K}$, and at $\SI{20}{\milli\K}$, measured from the $I-V$ curves. The characteristic voltage $I\ped c R\ped N$ decreases by increasing the barrier thickness, as well as the corresponding Josephson frequency $\omega\ped c=I\ped cR\ped N2e/\hbar$. At $\SI{4.2}{\K}$, it ranges from $\SI{80}{\giga\hertz}$ for the thinnest junction to $\SI{1}{\giga\hertz}$ for the thickest one. At lower temperatures, we measure higher values of the $I\ped cR\ped N$ product. These values are higher than those usually achieved for SFS JJs and comparable to those of some SIFS heterostructures~\cite{Bergeret2005,Buzdin2005,Weides2006,Bannykh2009,Khaire2010,Larkin2012}. For barrier thicknesses lower than $\SI{2.5}{\nm}$, the characteristic voltage is as high as a few millivolts~\cite{Senapati2011,Massarotti2015,Caruso2019}.
\begin{table*}
	\centering
	\caption[]{Parameters of the measured spin-filter junctions: thickness $t$, subgap resistance $R\ped{sg}$, quality factor $Q_0$ and capacitance $C$ calculated with Eq.~\ref{Eq:3}. $R\ped{sg}$ and $Q_0$ have been determined by fitting the $I-V$ curves according to the TJM model. The errors on the subgap resistance and the quality factor are of the order of $10\%$ and $6\%$, respectively, while the error on the capacitance ($20\%$) is obtained by propagation of the errors on $Q_0$ and $R\ped{sg}$, and are of the same order of magnitudes of those in Ref.~\cite{Martinis1987}.}
	\begin{ruledtabular}
		\begin{tabular}{cccccc}
			$t\;(\si{\nm})$ &  $R\ped{sg}@\SI{4.2}{\K}\;(\si{\ohm})$ & $R\ped{sg}@\SI{300}{\milli\K}\;(\si{\ohm})$ & $Q_0@\SI{4.2}{\K}$ & $Q_0@\SI{300}{\milli\K}$ & $C$ ($\si{\pico\farad}$) \\
			\hline
			$2.5$ & $59$ & $93$ &$16$ & $48$ & $1.6\pm0.3$ \\
			$3.0$ & $82$ & $350$ & $7.3 $ & $35$ & $1.1\pm0.2$  \\
			$3.5$ & $440$ & $1700$ & $6.6$ & $32$ & $0.26\pm0.05$\\
			$4.0$ & $3000$ & $13000$ & $2.6$ & $26$ & $0.018\pm0.003$ \\
		\end{tabular}
	\end{ruledtabular}
	\label{Tab:1}
\end{table*}
\begin{figure}
	\centering
	\includegraphics[width=8.6cm]{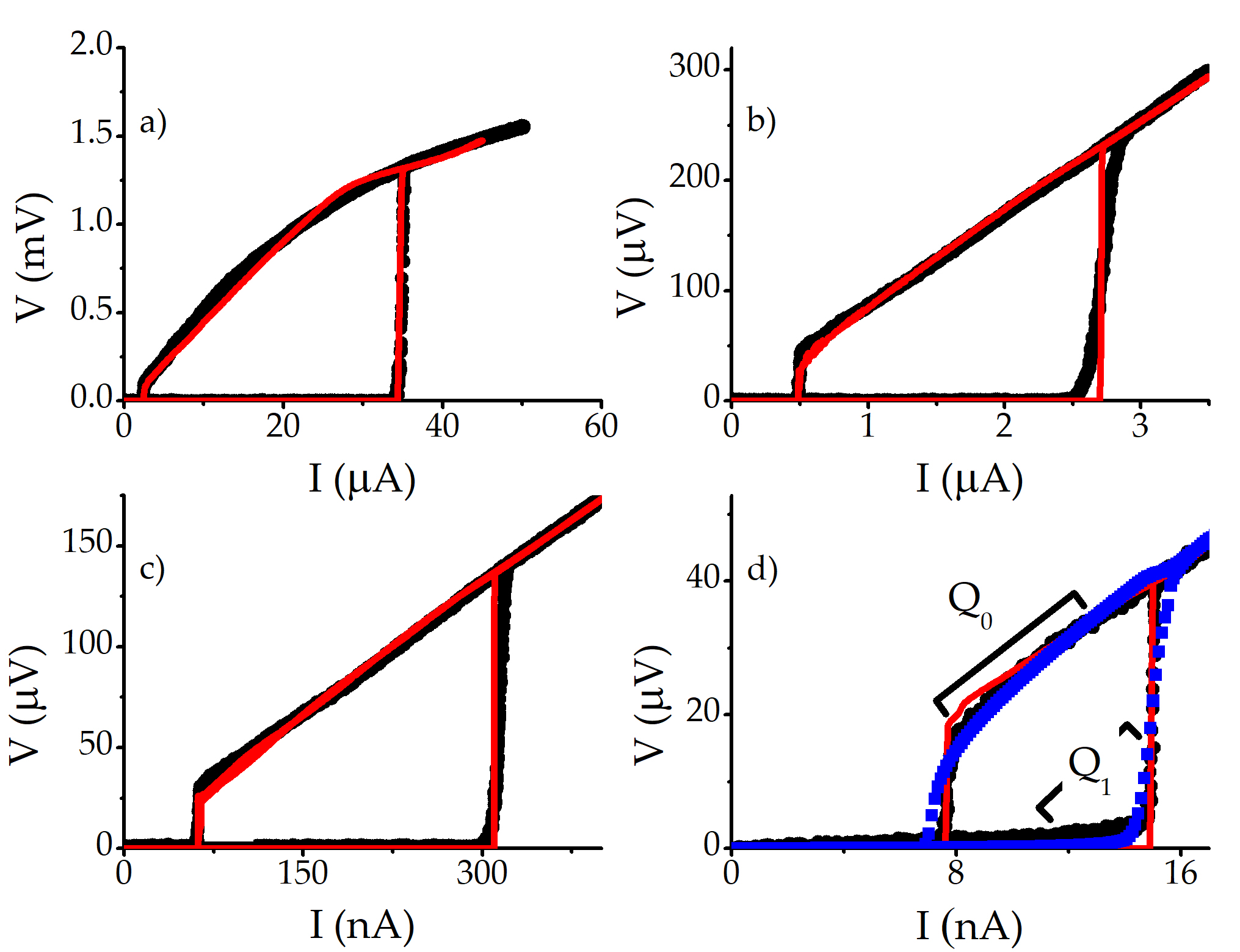}
	\caption[]{Measured $I-V$ curves at $\SI{4.2}{\K}$  (black points) and TJM model simulation by using \textsc{pscan2}\\
		 software (red curve) for high spin-filter JJs with thicknesses $t$: a) $\SI{2.5}{\nm}$, b) $\SI{3.0}{\nm}$, c) $\SI{3.5}{\nm}$, d) $\SI{4.0}{\nm}$. Quality factor $Q_0$ and subgap resistance $R\ped{sg}$ estimated from the simulations are collected in Tab.~\ref{Tab:1}. The blue squares in d) represent the frequency dependent RCSJ model fit curve, obtained for $Q_0=2.8$ and $Q_1=0.13$.}
	\label{Fig:2}
\end{figure} 

In Fig.~\ref{Fig:2}, we present the $I-V$ curves measured at $\SI{4.2}{\K}$ (black points) and TJM simulations (red straight lines) obtained by using \textsc{pscan2}. We collect in Tab.~\ref{Tab:1} the fitting parameters $R\ped{sg}$ and $Q_0$.  The thicker the barrier is, the higher is the subgap resistance~\cite{53,54,56}. The low-frequency quality factor $Q_0$ decreases with the thickness. This is due to both the decrease of $I\ped c$ and of $C$ of the barrier with the thickness~\cite{Barone1982}.  
\begin{figure}[t]
	\includegraphics[width=8.6cm]{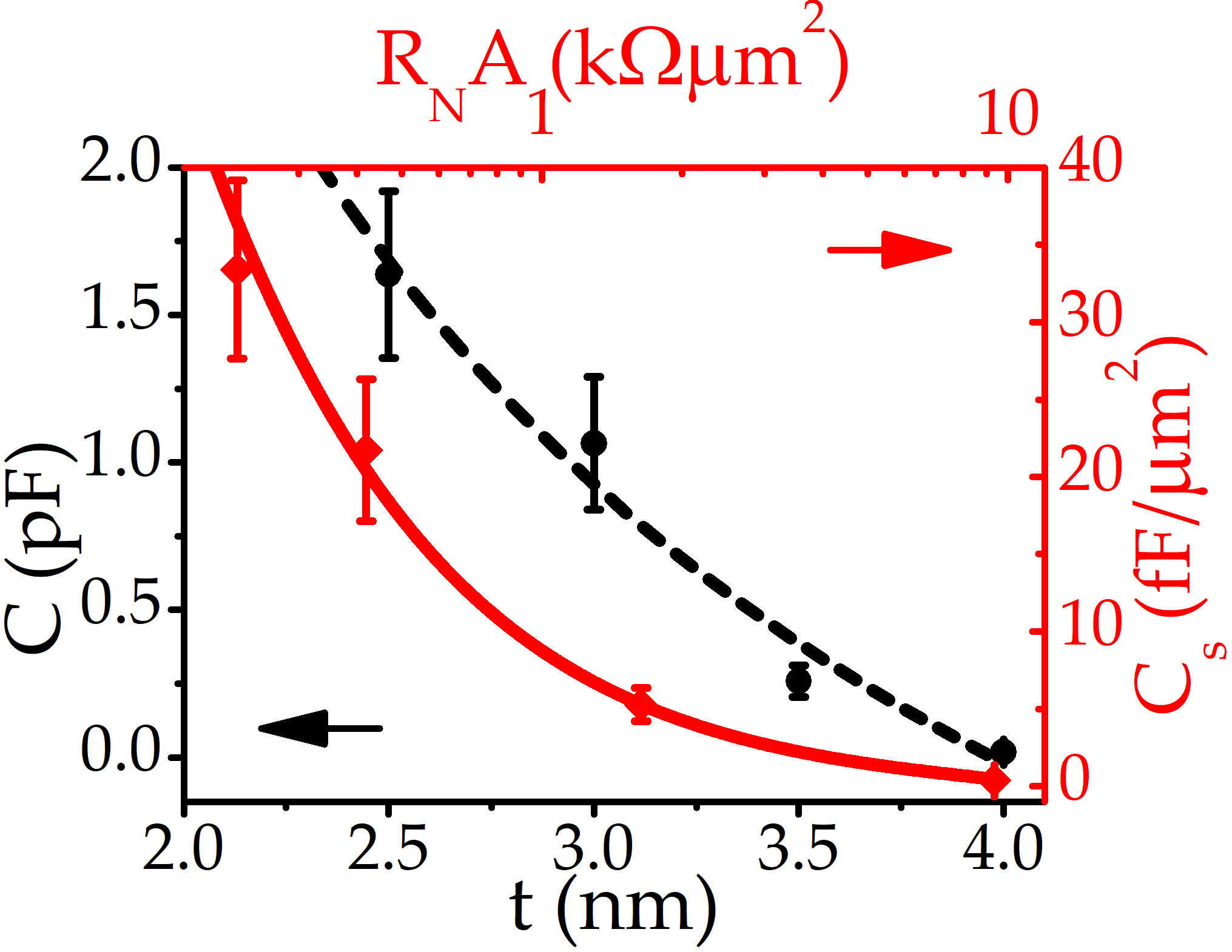}
	\caption[]{In black: capacitance values of spin-filter JJs as a function of the \ce{GdN} barrier thickness $t$ (black circles), along with parallel-plate capacitance $C(t)$ fit (dashed curve). In red: specific capacitance $C\ped s$ of the analyzed junctions as a function of $R\ped NA$ (red diamonds) along with a tunnel barrier model fit (straight line, see Eq.~\ref{Eq:2}). The error bars on $C$ and $C\ped s$ are calculated using the propagation of the errors on $R\ped{sg}$, $Q_0$ and $I\ped c$.}
	\label{Fig:3}
\end{figure} 

The decrease in the $Q_0$ factor  for increasing $t$ indicates a smooth transition from an underdamped regime ($Q_0\sim10$) to a moderately damped regime with phase diffusion (PD) ($Q_0\sim 1$)~\cite{Kautz1990,Massarotti2012,Stornaiuolo2013}. The presence of the PD regime is confirmed by the finite slope in the supercurrent branch for the junction with a  $\SI{4.0}{\nm}$-thick barrier~\cite{Kautz1990}, which \textsc{pscan2} simulations cannot reproduce, since they do not take PD processes into account. Monte Carlo simulations can reproduce the finite slope in the supercurrent branch, taking into account multiple escape and retrapping processes in the phase dynamics, which are particularly relevant for low values of the $Q_1$ factor and $E\ped J$ comparable with the thermal energy $k\ped BT$, as in the case of the spin-filter junction with a $\SI{4.0}{\nm}$-thick barrier~\cite{Kautz1990}. In Fig.~\ref{Fig:2}~(d), a Monte Carlo fit according to the frequency dependent RCSJ model is shown (blue square points), with high-frequency $Q_1 = 0.13$ and low-frequency $Q_0 = 2.8$. This is consistent with the outcomes based on the TJM model. 

The environment plays an important role in determining the value of $Q_1$. The ratio between the low- and high-frequency quality factors $Q_1/Q_0$ equals the ratio between the resistance of the environment $R\ped{env}$ and the subgap resistance, $R\ped{env}/R\ped{sg}$, since $Q_0$ is written in terms of the quasiparticle dissipation (Eq.~\ref{Eq:3}), while $Q_1$ can be expressed in terms of the environment resistance $R\ped{env}$~\cite{Martinis1987,Martinis1989,Kautz1990}. For the junction with a \ce{GdN} barrier thickness of $\SI{4.0}{\nm}$, $R\ped{env}$ is approximately $\SI{150}{\ohm}$, which is of the same order of magnitude of the resistance of the lines in our experimental setup.

Our analysis allows us to estimate the capacitance $C$ of the barrier and its dependence on the barrier thickness, using Eq.~\ref{Eq:3} (see Tab.~\ref{Tab:1}). The value of $C$ for the thinnest \ce{GdN} barrier is consistent with a previous estimation based on SCDs measurements~\cite{Massarotti2015}. In Fig.~\ref{Fig:3}, we plot  the junction capacitance $C$ as a function of the \ce{GdN} barrier thickness $t$ (black circle points) and the fitting function for the capacitance in a parallel-plate capacitor $C=\epsilon_0\epsilon\ped r A/t$ (black dashed line), where $\epsilon_0=\SI{8.85}{\pico\farad/m}$ is the vacuum dielectric permittivity and $\epsilon\ped r$ is the \ce{GdN} relative permittivity, which acts as a fitting parameter. The estimated $\epsilon\ped r=(20\pm8)$ is consistent with the \ce{GdN} permittivity $\epsilon\ped r=26.5$ obtained with spectroscopic measurements on isolated \ce{GdN} thin films~\cite{63}, providing an additional validation of the fitting procedure. The $R\ped NA$ product as a function of the specific capacitance $C\ped s=C/A$ (red diamonds in Fig.~\ref{Fig:3}), follows the expected behavior for tunnel JJs~\cite{Kawakami2003}. The red line in Fig.~\ref{Fig:3} is the function
\begin{equation}
R\ped NA(C\ped s)=\frac{2A\epsilon_0\epsilon\ped r}{3C\ped s\sqrt{4m\ped e\bar{E}}}\left(h/e\right)^2e^{\frac{2\epsilon_0\epsilon\ped r}{\hbar C\ped s}\sqrt{4m\ped e \bar{E}}},
\label{Eq:2}
\end{equation} 
which is obtained by replacing $t$ in equation~\ref{Eq:1} with its dependence on the specific capacitance $C\ped s$, $t=\epsilon_0\epsilon\ped r/C\ped s$.

\begin{figure}
	\centering
	\includegraphics[width=8.6cm]{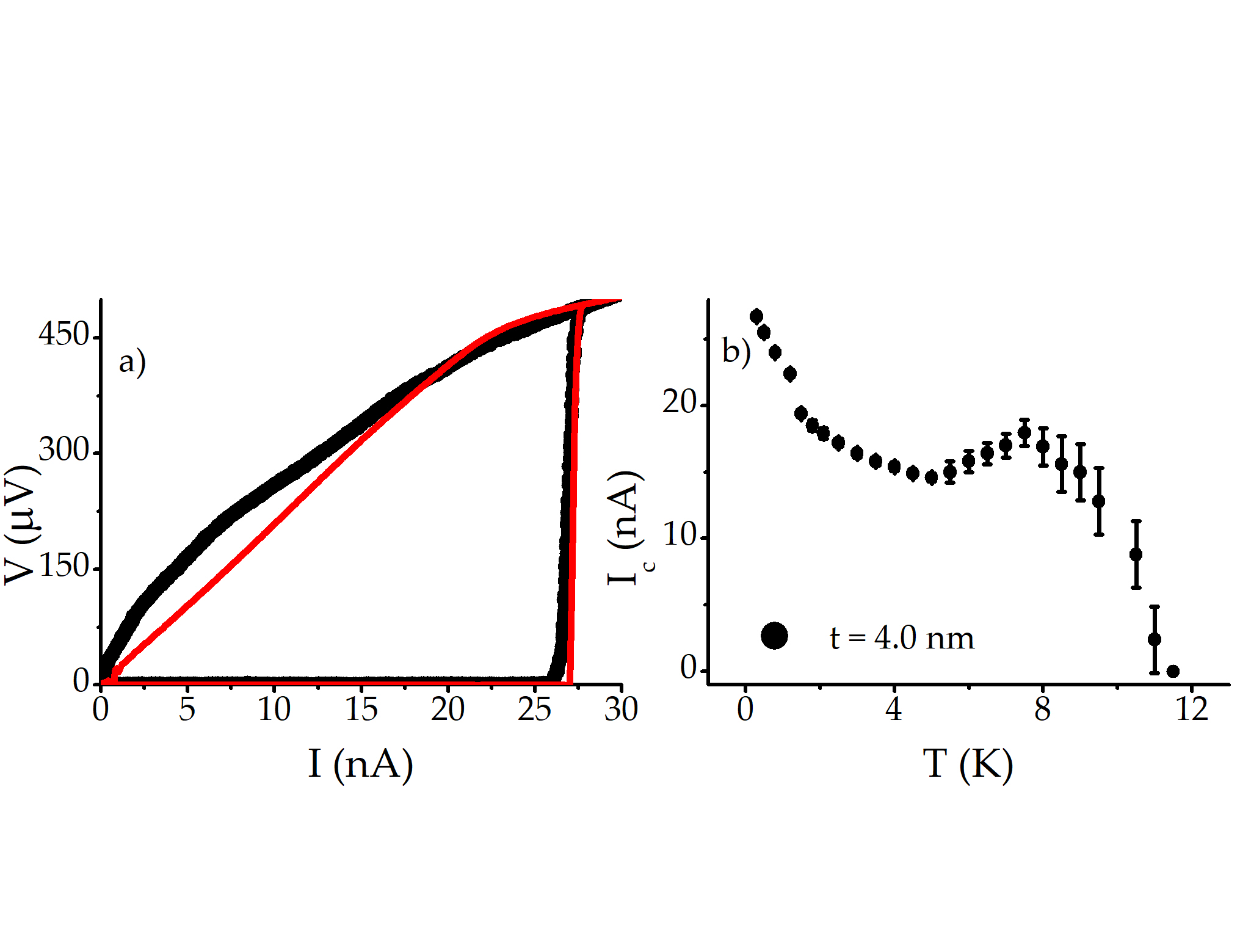}
	\caption[]{In a): measured $I-V$ characteristic at $\SI{300}{\milli\K}$ (black points) and TJM model simulation by using \textsc{pscan2} software (red straight line) for the spin-filter JJ with $t=\SI{4.0}{\nm}$. Quality factor $Q_0$ and subgap resistance $R\ped{sg}$ estimated from the simulations are collected in Tab.~\ref{Tab:1}. In b): incipient $0$-$\pi$ transition in the $I\ped c(T)$ for the spin-filter JJ with $t=\SI{4.0}{\nm}$, as reported in Ref.~\cite{Caruso2019}.}
	\label{Fig:7}
\end{figure}

In Fig.~\ref{Fig:7}~(a) we show the $I-V$ characteristic measured at $\SI{300}{\milli\K}$ (black points) and TJM simulations (red straight lines) for the junction with a $\SI{4.0}{\nm}$-thick barrier, which corresponds to the highest spin-filtering efficiency analyzed in this work. $Q_0$ and $R\ped{sg}$ for all the devices are collected in Tab.~\ref{Tab:1}. The best-fit curve at $\SI{300}{\milli\K}$ is characterized by a smaller $R\ped{sg}$ compared to the experimental one. We can attribute this deviation to the unconventional magnetic activity discussed in Ref.~\cite{Caruso2019}, which is at a maximum in the case of most spin-polarized JJs, where the magnetic nature of the barriers manifests in a steep increase of $I\ped c(T)$ below $\SI{2}{\K}$ [Fig.~\ref{Fig:7}~(b)]~\cite{Caruso2019}. The conventional TJM model does not take the magnetic activity in the $I\ped f$ barrier into account, nor the spin-dependent tunneling mechanism and the unconventional thermal behavior of $I\ped c$, thus giving a systematic underestimation of $R\ped{sg}$, as shown in Fig.~\ref{Fig:7}~(a). However, despite the presence of these deviations, $R\ped{sg}$ estimated for all the junctions increases when decreasing the temperature $T$ due to the tunnel nature of the conduction mechanisms in the system~\cite{54} and $Q_0$ increases because of the increase of $R\ped{sg}$, as expected.

\begin{figure}
	\centering
	\includegraphics[width=8.6cm]{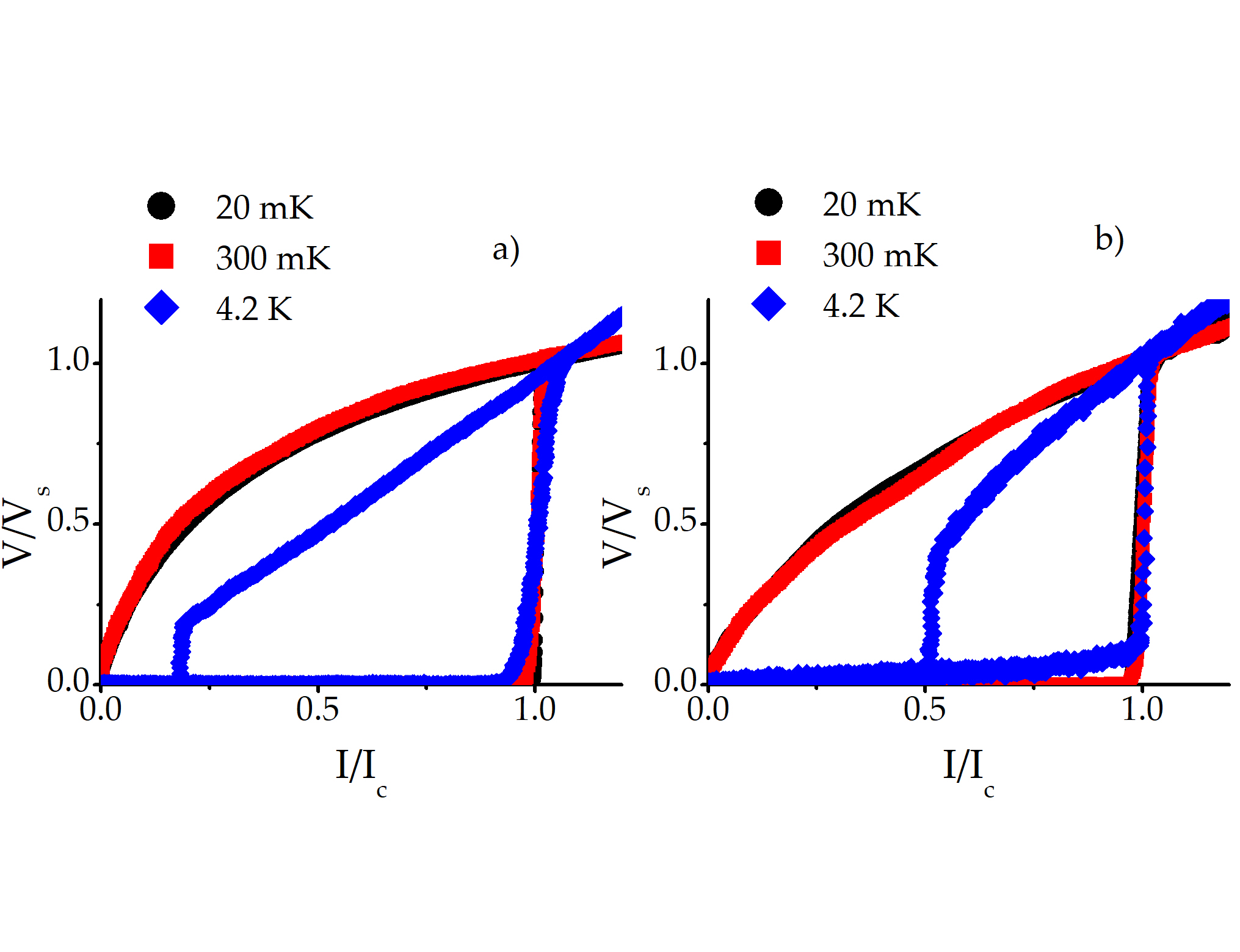}
	\caption[]{Normalized $I-V$ curves at $\SI{20}{\milli\K}$ (black points), $\SI{300}{\milli\K}$ (red squares) and $\SI{4.2}{\K}$ (blue diamonds) for spin-filter JJs with thicknesses $t$: a) $\SI{3.0}{\nm}$ and b) $\SI{4.0}{\nm}$. The current is normalized to the critical current $I\ped c$, while the voltage is normalized to the switching value $V\ped s$.}
	\label{Fig:9}
\end{figure}

In Fig.~\ref{Fig:9}, we finally present a comparison between the normalized $I-V$ curves measured at $\SI{4.2}{\K}$ and $\SI{300}{\milli\K}$ and the $I-V$ characteristics at $\SI{20}{\milli\K}$ for two of the junctions with the highest $P$: in a) $t=\SI{3.0}{\nm}$ and $P=93\%$ and b) $t=\SI{4.0}{\nm}$ and $P=98\%$. The current is normalized to $I\ped c$, while the voltage is normalized to the switching value $V\ped s$ to compare the subgap branches of the $I-V$ curves. $I\ped c$ at $\SI{20}{\milli\K}$ are $\SI{4.75}{\micro\ampere}$ for a) and $\SI{29}{\nano\ampere}$ for b). The critical current at $\SI{300}{\milli\K}$ is $\SI{4.64}{\micro\ampere}$ for a) and  $\SI{26.7}{\nano\ampere}$ for b). The amplitude of the hysteresis in the $I-V$ curves increases when going toward lower temperatures, pointing to an increase of $Q_0$ and also as a consequence of $R\ped{sg}$.

\section{Discussions and concluding remarks}
\label{Discussions}

The use of the TJM model on parent compounds allows to achieve a consistent and robust picture of tunnel-ferromagnetic JJs with a quantitative insight on key electrodynamic parameters, such as $Q_0$, $R\ped{sg}$ and $C$.

The estimated $Q_0$ values at $\SI{4.2}{\K}$ are up to two orders of magnitude higher compared to those of standard SFS heterostructures that typically operate in the overdamped regime like SNS JJs, with $\beta$ ranging from $10^{-3}$ to $10^{-1}$~\cite{Bulaevskii1977,Barone1982,Likharev1986}. $Q_0$ values are of the same order of magnitude of conventional SIS junctions commonly used to drive and for the read-out of components in quantum and classical circuits~\cite{Castellano2006,Shcherbakova2015}. Moreover, the $Q_0$ values increase up to one order of magnitude for the $\SI{4.0}{\nm}$ thick barrier, when lowering $T$ to $\SI{300}{\milli\K}$. This sets a lower limit that can only increase at lower temperatures (see Fig.~\ref{Fig:9}), and suggests possible implementation of spin-filter JJs in low-dissipative $\pi$-qubits. $\pi$-superconducting RF-SQUIDS with ferromagnetic-insulating barriers were only theoretically suggested as \emph{quiet} qubits efficiently decoupled from the fluctuations of an external magnetic field~\cite{Kawabata2006,Kawabata2010}. A spin-filter JJ with $t=\SI{3.0}{\nm}$ and an $A\sim\SI{50}{\micro\m^2}$ has an estimated charging energy $E\ped c\sim\SI{900}{\micro\K}$, and a Josephson energy at $\SI{20}{\milli\K}$ $E\ped J\sim\SI{100}{\K}$, which means $E\ped J/E\ped c\sim10^5$, suitable for a flux-qubit~\cite{Devoret2013,Krantz2019}. 

In the frame of the $0$-$\pi$ technology, spin filter JJs analyzed in this work can be implemented also as complementary $\pi$-junctions for phase-bias of conventional flux-qubit (passive elements) in which the high values of the subgap resistance $R\ped{sg}$ could increase the dephasing time of the overall circuit~\cite{Kato2007}. The dephasing time is proportional to $E\ped J^2R\ped {sg}$ of the $\pi$-junction~\cite{Kato2007}. The subgap resistance in this work ranges from tens of ohms to some kilo-ohms at $\SI{4.2}{\K}$, but when decreasing $T$, $R\ped{sg}$ increases from a factor $2$ to $5$ increasing $t$ at $\SI{300}{\milli\K}$. The dephasing time of a circuit with a spin-filter JJ with a $t=\SI{3.0}{\nm}$ thick barrier can be comparable with that of circuits with SIFS $\pi$-junctions~\cite{Weides2006}, and can increase of at least a factor $100$ compared to circuits with standard metallic $\pi$ shifters~\cite{Feofanov2010,Shcherbakova2015}. In standard metallic SFS JJs typical resistances are at most $\sim\SI{1}{\ohm}$, while $R\ped{sg}$ for the junction with a $\SI{3.0}{\nm}$ thick barrier at dilution temperature is at least $\SI{350}{\ohm}$.

The subgap resistance is crucial for the engineering of transmon qubits. As suggested in Ref.~\cite{Serniak2018}, in these circuits quasiparticle tunneling can affect the relaxation and coherence times~\cite{Serniak2018}. The values obtained in this work can be promising even for potential application of tunnel-ferromagnetic JJs in transmon qubits. The order of magnitude of the ratio $E\ped J/E\ped c$ for the investigated junctions scales with the thickness from $10^6$ to $10$. Adapting the area of the devices to conventional dimensions in transmon qubits ($A\sim \SI{1}{\micro\m^2}$), lower values of $E\ped J/E\ped c$ can be achieved, falling in the typical range of transmon qubit~\cite{11,Devoret2013,Krantz2019}. As an example, reducing the cross section to $A\sim\SI{1}{\micro\m^2}$, $E\ped J$ of the spin-filter JJ with $t=\SI{3.5}{\nm}$ becomes $\sim\SI{280}{\milli\K}$, while $E\ped c$ becomes $\sim\SI{180}{\milli\K}$, so that $E\ped J/E\ped c\sim2$. Moreover, reducing the junction area by a factor $\sim 50$, $R\ped{sg}$ should increase up to values of the order of $50-\SI{100}{\kilo\ohm}$, thus further reducing quasiparticle noise. The same arguments are valid for the junction with $t=\SI{4.0}{\nm}$ \ce{GdN} barrier, which is characterized by a subgap resistance $\sim 10$ times higher. 

In conclusion, this work represents the first electrodynamic characterization of spin-filter JJs, and a fundamental step to use these devices as active elements in superconducting circuits.  Our comparative and self-consistent approach allows to obtain the scaling-law as a function of the barrier thickness of fundamental electrodynamic parameters, such as $C\ped s (t)$, $R\ped{sg}(t)$ and $Q_0(t)$, providing the possibility to engineer spin-filter JJs as a function of the junction area in order to meet specific circuit requirements. Even if the ferromagnetic JJs analyzed in this work are not ideal SIS JJs, we succeeded in the determination of these fundamental electrodynamic parameters at $\SI{4.2}{\K}$ by using a conventional TJM model,  and we provided a lower bound for $R\ped{sg}$ and $Q_0$ at $\SI{300}{\milli\K}$. The underestimation of $R\ped{sg}(t)$ observed at $\SI{300}{\milli\K}$ is due to the absence of the spin-filtering effect and of the magnetic activity of the barrier in the TJM model. Further studies are needed to implement a microscopic modelization of peculiar properties of the $\text{I}\ped f$ barrier, such as spin-selective tunneling mechanisms and triplet correlations.

The same approach can be successfully extended to different types of tunnel junctions other than conventional SIS JJs, for instance, multilayered SIFS JJs, and can provide the possibility to engineer special circuits other than conventional flux and transmon qubits, in which ferromagnetic-tunnel junctions can be tuned by external microwaves and are capacitively coupled to standard superconducting circuits~\cite{20,21}.

\begin{acknowledgments}
	The authors thank G. Campagnano, S. Poletto, A. Miano and A. Kirichenko for fruitful discussions. H.G.A., R.C., D.M. and F.T. also thank NANOCOHYBRI project (COST Action CA 16218).
\end{acknowledgments}

\bibliography{refTJM}

\appendix

\section{The spin-filtering effect}
\label{Appendix2}

Josephson junctions with \ce{GdN} barriers show a spin-filtering effect due to the simultaneous presence of tunnel conduction mechanisms and a magnetic exchange field $h$ in the ferromagnetic phase of the barrier. When the \ce{GdN} becomes ferromagnetic ($T\ped{Curie}\sim\SI{40}{\K}$), the presence of exchange interactions leads to a spin selectivity of the tunneling processes: spin up (down) will see different barrier heights $E_{\uparrow(\downarrow)}=E_0\mp h/2$, with $E_0$ energy barrier height in the paramagnetic phase of the \ce{GdN}. Carriers that relive a higher barrier will be filtered out, thus giving a net spin-polarized current~\cite{Senapati2011}. 
\begin{figure}[b]
	\centering
	\includegraphics[width=8.6cm]{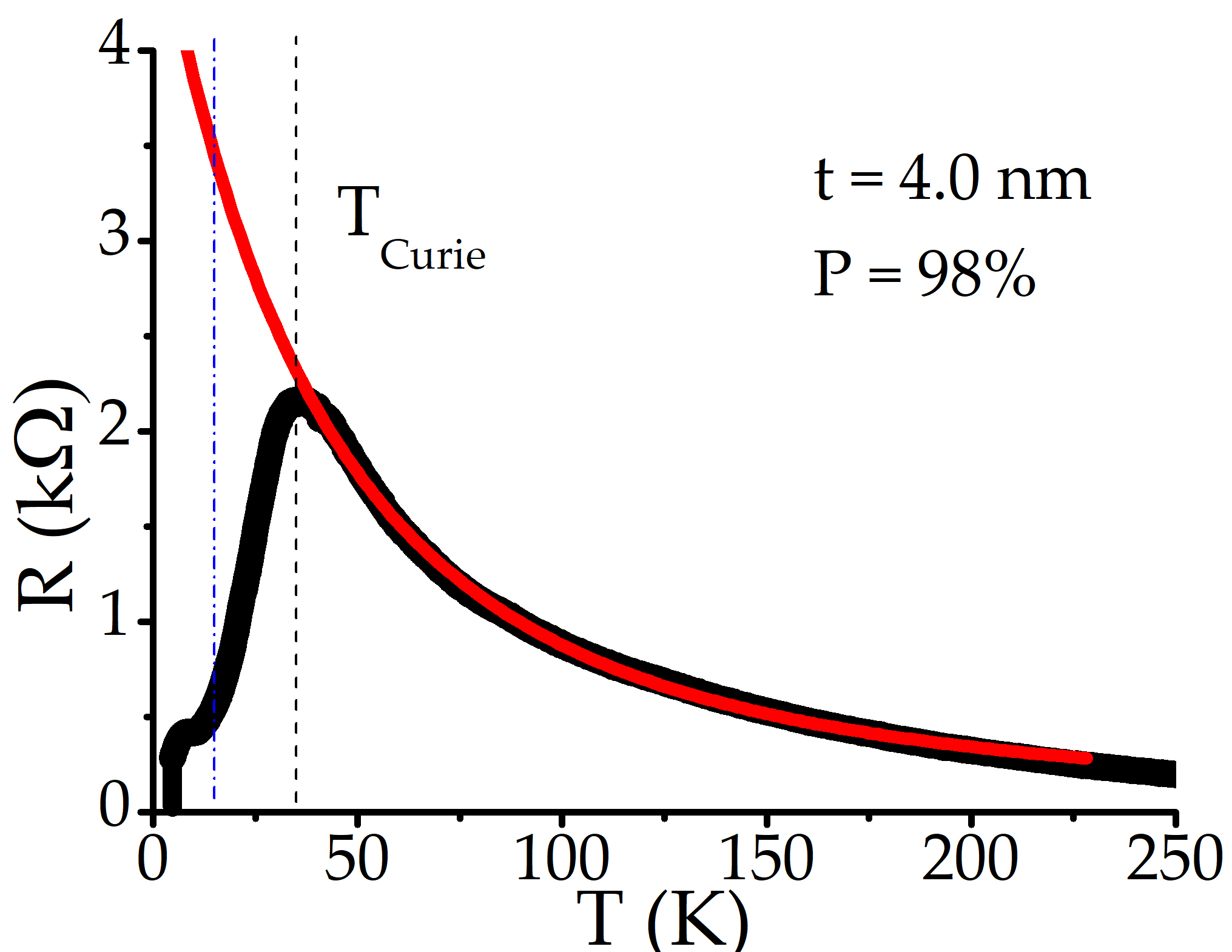}
	\caption[]{Measured $R(T)$ curves (black points) and semiconducting model fit (red straight line) above the Curie temperature $T\ped{Curie}$ of the device  (dashed black line) for the junction with $t=\SI{4.0}{\nm}$. The dash dotted blue line indicates the temperature $\SI{15}{\K}$ at which we calculate the $P$.}
	\label{Fig:6}
\end{figure} 

An experimental measure of the spin-filtering efficiency is obtained from the $R(T)$ curves, since it can be defined as 
\begin{equation}
P=\left|\frac{\sigma_{\uparrow}-\sigma_{\downarrow}}{\sigma_{\uparrow}+\sigma_{\downarrow}}\right|,
\end{equation}
where $\sigma_{\uparrow(\downarrow)}$ is the tunnel conductance through the barriers seen by spin up (down) carriers. In the limit of small magnetic exchange fields, the spin-filtering efficiency reduces to 
\begin{equation}
P\sim\tanh\left(\coth^{-1}\left(\frac{R^{*}}{R}\right)\right),
\end{equation}
where $R$ and $R^{*}$ correspond to the measured resistance and the resistance in the absence of magnetic exchange field, respectively~\cite{Senapati2011}.
\begin{figure}[b]
	\centering
	\includegraphics[width=8.6cm]{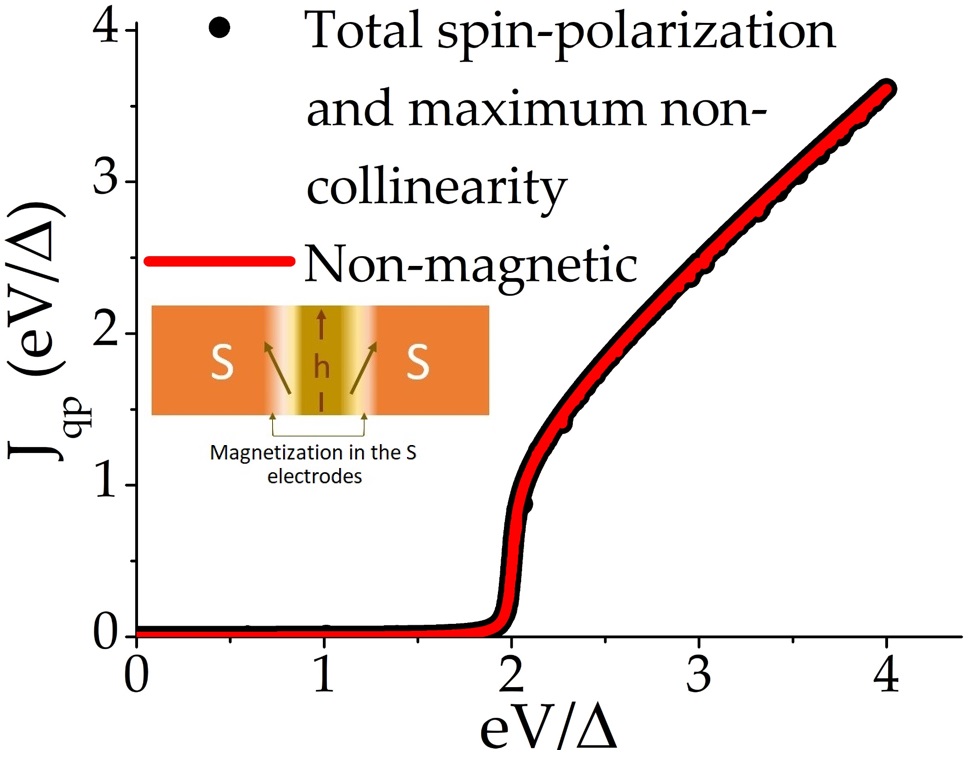}
	\caption[]{Comparison between the normalized subgap branch $J\ped{qp}=I\ped{qp}R\ped N/V$ for a non-magnetic tunnel JJ (red straight line) and a perfect spin-filter tunnel JJ (black points). The expression of the quasiparticle current used for the simulations can be found in~\cite{Bergeret2012}. The parameters used to reproduce the curves are: $P=0$, $h=0$, $\Delta=1$, $\eta=0.01\Delta$ (damping factor) and $T=\SI{4.2}{\K}$ for the non-magnetic junction and $P=1$, $h=0.4\Delta$, $\Delta=1$, $\eta=0.01\Delta$ for the spin-filter JJ. The angles $\alpha$ and $\beta$ between $h$ and the magnetization induced in the superconducting electrodes are $\alpha=\beta=\pi/2$.}
	\label{Fig:8}
\end{figure}

In Fig.~\ref{Fig:6} we show the $R(T)$ curve (black points) for the spin-filter JJ with a \ce{GdN} barrier thickness of  $\SI{4.0}{\nm}$. The red straight line is the semiconducting fit performed in the paramagnetic phase of the barrier. This curve allows to obtain the resistance in the absence of magnetic exchange field $R^{*}$. Below $T\ped{Curie}$ (dashed black line), we can observe a decrease in the resistance because one spin channel is favored in the conduction. 

The spin-selective tunneling processes affect the normal resistance $R\ped N$ too, which is defined as $R\ped N=1/(4\pi (eN(0))^2(\sigma_{\uparrow}+\sigma_{\downarrow}))$, with $N(0)$ density of state at the Fermi level~\cite{Bergeret2012}. 

The magnetic exchange field $h$ in the barrier induces a magnetization in the superconducting electrodes of spin-filter JJs~\cite{Bergeret2012}. The angles between the magnetization in the electrodes and the magnetic exchange field will be denoted as $\alpha$ and $\beta$~\cite{Bergeret2012}. Tunneling of spin-polarized carriers appears only if the angles $\alpha$ and $\beta$ between $h$ and the magnetization in the left and right superconducting electrodes, respectively, are different from $0$ and $\pi$ [Fig.~\ref{Fig:8}]~\cite{Bergeret2012}. The quasiparticle current in non-magnetic devices ($P=0$, $h=0$), and in magnetic JJs with total spin polarization ($P=1$, $h<\Delta$, being  $\Delta$ the superconducting gap of the electrodes) and maximum non-collinearity between $h$ and the magnetization in the superconducting electrodes ($\alpha=\beta=\pi/2$) has the same analytic expression, and the $I-V$ curves are comparable [Fig.~\ref{Fig:8}]. We verify this statement using the expression for the quasiparticle current in spin-filter JJs proposed in~\cite{Bergeret2012}. This result justifies the use of a conventional TJM model, in which there is no explicit introduction of a magnetic exchange field in the barrier, when fitting the $I-V$ curves in the ideal case of perfect spin polarization.
\end{document}